\documentclass[submission]{eptcs}

\providecommand{\event}{GCM 2022} 
\usepackage{breakurl}             
\usepackage{amsthm}
\usepackage{amssymb}
\usepackage{amsmath}
\usepackage{mathtools}
\usepackage{oplotsymbl}
\usepackage{MnSymbol}
\usepackage{trimclip}
\usepackage{graphicx}
\usepackage{stmaryrd}

\makeatletter
\newcommand*{\twoheadrightsquigarrow}{\rightsquigarrow\joinrel\mathrel{\mathpalette\@twoheadrightsquigarrow\relax}}
\newcommand*{\@twoheadrightsquigarrow}[2]{%
   \clipbox{{.7\width} 0pt 0pt {-.2\height}}{$\m@th#1\rightsquigarrow$}%
}

\makeatletter
\newcommand*{\mydiv}{%
  \mathbin{%
    \mathpalette\@mydiv{}%
  }%
}
\newcommand*{\@mydiv}[2]{%
  \sbox0{$#1\vcenter{}$}%
  \sbox2{\scalebox{1.15}{$#1.\m@th$}}%
  \sbox0{%
    \raisebox{-.25\ht2}{\rlap{\copy2}}%
    \raisebox{\dimexpr2\ht0-\ht2+.25\ht2\relax}{\copy2}%
  }%
  \sbox2{$#1\sim$}%
  \rlap{\copy2}%
  \hbox to \wd2{\hfil\copy0\hfil}%
}
\makeatother

\newtheorem{definition}{Definition}

\title{A Foundation for Functional Graph Programs:\\ The \textit{Graph Transformation Control Algebra} (GTA)}
\author{Jens H. Weber
\institute{Department of Computer Science \\ University of Victoria \\ Victoria, Canada, BC}
\email{jens@acm.org}
}
\def\titlerunning{A Foundation for Functional Graph Programs}
\def\authorrunning{J. H. Weber}
\begin{document}
\maketitle

\begin{abstract}
Applications of graph transformation (GT) systems often require control structures that can be used to direct GT processes. Most existing GT tools follow a stateful computational model, where a single graph is repeatedly modified \textit{in-place} when GT rules are applied. The implementation of control structures in such tools is not trivial. Common challenges include dealing with the non-determinism inherent to rule application and transactional constraints when executing compositions of GTs, in particular atomicity and isolation. The complexity of associated transaction mechanisms and rule application search algorithms (e.g., backtracking) complicates the definition of a formal foundation for these control structures. Compared to these stateful approaches, {\em functional} graph rewriting presents a simpler (stateless) computational model, which simplifies the definition of a formal basis for (functional) GT control structures. In this paper, we propose the \textit{Graph Transformation control Algebra} (GTA) as such a foundation. The GTA has been used as the formal basis for implementing the control structures in the (functional) GT tool \emph{GrapeVine}. 

\end{abstract}

\section{Introduction}

There are diverse practical applications of graph transformations (GT) in software engineering, computer science and beyond~\cite{GT20-GraphTransformation}. They often require or benefit from {\em control structures} that can be used to restrict and direct the application of GTs. Heckel and Taentzer categorize four approaches for controlling the application of GTs, including non-terminals, dedicated control expressions, integrity constraints, and procedural abstraction~\cite{HT20-BeyondIndividual}. Real-world applications often employ a combination of these approaches. Most existing GT tools follow a stateful model of computation where rule applications and control expressions mutate the program state. Multiple rule applications may need to be grouped into \textit{atomic} units that are executed entirely or not at all and in isolation of other rule applications. Finding solutions with non-deterministic rule applications may require search mechanisms like backtracking and heuristic optimization. Implementing such sophisticated mechanisms is not trivial and their semantics is often not formalized, which impedes formal reasoning about the behaviour of non-deterministic GT programs.

It is well known that stateless computation provides for a simpler semantic model when compared to stateful computation. We recently presented a functional GT tool (called \emph{GrapeVine}) that implements a stateless computational model with deterministic operators for rule application~\cite{Web22-ToolSupport}. This paper presents the formal foundation for the control structures provided in this tool. We use an algebraic approach to define a set of operations referred to as the {\em Graph Transformation control Algebra} (GTA). In addition to providing a precise semantics for the control structures provided by \textit{GrapeVine}, we hope that the GTA will enable future research on  analysis and optimization of functional, deterministic GT programs, analogous to the use of the relational algebra for database query analysis and optimization.

The rest of this paper is structured as follows. We start with a  brief introduction to graphs, graph transformations and graph constraints.  Sec.~\ref{sec-related} provides a description of related work on control structures for GT programs.  We introduce the \textit{ Graph Transformation control Algebra} (GTA) and our notion of programmed graph transformation systems in Sec.~\ref{sec:gta}. Sec.~\ref{sec-impl} provides a description of an implementation of the control structures in the tool \textit{GrapeVine}, which are based on the GTA. Finally, we offer conclusions and discuss current work in Sec.~\ref{sec-conc}.

\section{Preliminaries}
\label{sec-prelim}

For the purpose of this paper, we limit ourselves to a definition of basic forms of graphs and graph transformations (GTs). While many GT tools (including \textit{GrapeVine}) provide more advanced concepts (e.g., labels, attributes, typed graphs), the definitions below can easily be extended accordingly. The reader is referred to \cite{FunGT06} for a more complete introduction to graph transformations.

\begin{definition}[Graph]
A \textit{graph} is a tuple $G:(N,E,s,t)$ where $N$ is a finite set of \emph{nodes}, $E$ is a finite set of \emph{edges}, and $s,t:E \rightarrow N$ are total \emph{source} and \emph{target} functions, respectively. 
\end{definition}

\begin{definition}[Rule]
\color{black}
A (GT) \emph{rule} is defined as a pair of graphs $(L,R)$ that share a common (possibly empty) \textit{interface} graph $I = L \cap R $ containing all nodes and edges shared between in L and R. $L$ is called the rule's \emph{left-hand side} and $R$ is called its \emph{right-hand side}.
\color{black}
\end{definition}

\begin{definition}[Transformation]
An application of a rule  \color{black} $r:(L,R)$ \color{black} to a given host graph $G$ requires a {\em match} of the left-hand side in $G$, i.e., the existence of a graph morphism $L \xrightarrow{m} G$. The application deletes all elements from $G$ that are matched to elements that only appear on the rule's left-hand side ($L-I$),  creates elements for any elements that appear only on the rule's right-hand side ($R-I$), and embeds these new elements in the preserved context ($m(I)$). A more formal definition of a rule application based on category theory (double-pushout approach) is provided by Corradini et al. \cite{CMR+97-AlgebaricApproaches}. To avoid problems that may occur when deleting graph elements, a match has to satisfy the so-called \emph{gluing condition}, which consists of two parts: (1) the embedding context (attached edges) of any deleted nodes must also be identified and deleted by the rule (\emph{dangling condition}), and (2) deleted element must only have one pre-image in $L$ (\emph{identification condition}).

The transformation of a given graph $G$ into a graph $G'$ by applying rule $r$ at a valid match $m$ is denoted as  $G \overset{r,m}{\rightsquigarrow} G'$. \color{black} Note that in contrast to the definition in category theory (which identifies graphs up to isomorphism), we assume a transformation to produce a unique graph. \color{black}
\end{definition}

As mentioned before, we opted against adding more advanced concepts to our formalization of rules and transformations, e.g., (negative) application conditions and rule parameters. While \textit{GrapeVine} provides these concepts, they are not the focus of this paper. We do, however, introduce the notion of \textit{graph constraints},\cite{OEP08-ALogic} as  constraints are one of the mechanisms used to control the application of GT rules~\cite{HT20-BeyondIndividual}.

\begin{definition}[Atomic Constraint] 
An atomic (graph) constraint $I \xrightarrow{c} T$ is defined as a graph monomorphism between two graphs $I$ and $T$. A graph $G$ satisfies $c$ (denoted as $G \models c$) if for every monomorphism $I\xrightarrow{h} G$ there is a monomorphism $T\xrightarrow{f} G$ such that $h=f \circ c$. 
\end{definition}

\begin{definition}[Constraint]
A (graph) constraint  is inductively defined as either an atomic graph constraint  $c: I \rightarrow T$, a negation of a graph constraint (i.e., if $\kappa$ is a graph constraint, so is $\neg \kappa$), or the disjunction of two graph constraints (i.e., if $\kappa_1$ and $\kappa_2$ are graph constraints, so is $\kappa_1 \vee \kappa_2$).
\end{definition}

Now that we have introduced constraints, we extend our definition of graphs and transformations to constrained graphs and transformations on constraint graphs.

\begin{definition}[Constrained Graph]
A \textit{ constrained graph} is a tuple $(G,K)$ where $G$ is a graph and $K$ is a finite set of \textit{graph constraints} satisfied by $G$, i.e, $\forall \kappa \in K : G \models \kappa $. 
\end{definition}

\begin{definition}[Transformation of a constrained graph]
A transformation of a constrained graph  $(G,K) \overset{r,m}{\rightsquigarrow} (G',K)$ exists, if there exists a corresponding transformation for the unconstraint graph $G \overset{r,m}{\rightsquigarrow} G'$,  where the resulting graph satisfies  all constraints, i.e., $\forall \kappa \in K: G' \models \kappa$.
\end{definition}
 
 From here on, we only consider constrained graphs and transformations on constrained graphs. The reader should therefore assume that we refer to a ``constrained graph" whenever we use the term \textit{graph} without explicitly mentioning that we are talking about an unconstrained graph.  
 
A transformation $G \overset{r,m}{\rightsquigarrow} G'$ is also called a \textit{direct linear derivation}. In general, there are many possible matches for a rule in a given graph, i.e., multiple direct linear derivations are possible. We may just write  $G \overset{r}{\rightsquigarrow} G'$ to denote \textit{any} direct linear derivation, if we do not care about the match.

\section{Related Work}
\label{sec-related}

Most GT tools offer some constructs to control the application of GT rules. The inherent complexity related to stateful computation and non-deterministic choice between possible derivations complicates the implementation of execution engines for GT programs. To the best of our knowledge, \textit{PROGRES}~\cite{SWZ99-Progres}, \textit{Grape}~\cite{Web17-GrapeA}, and \textit{GP}~\cite{Plum09-TheGraph} are the only GT tools that implement backtracking mechanisms, capable of reconstructing graphs at non-deterministic choice points. The PROGRES project has been discontinued since its platform is no longer available. \textit{Grape} is still available on a modern platform but it is no longer under further development, in part due to the complexity of  maintaining the implemented backtracking system. Work on \textit{GP} appears to continue with the latest version of the language called \textit{GP2}. The execution of \textit{GP(2)} programs requires the York Abstract Machine (YAM) which handles a sophisticated data structure to keep track of non-determinism, choice points and environment frames.

\textit{PROGRES} and \textit{GP(2)} provide a formalized semantics for their control structures. The semantics definition for \textit{PROGRES} is complex and spans over three hundred pages. The formalization of \textit{GP(2)} is comparably simple. The original language (\textit{GP}) consist of only four types of commands: non-deterministic rule application, sequential composition, branching, and iteration~\cite{Plum09-TheGraph}. Its successor (\textit{GP2}) provides additional constructs, including a second kind of branching statement and an explicit operator for non-deterministic choice between sub-programs~\cite{Plum12-TheDesign}. Moreover, GP2 changed the semantics of control structures in order to allow for efficient implementation of branching and looping. In particular, failures in conditions of branching statements or loop bodies no longer enforce backtracking. 
While \textit{GP(2)} programs theoretically associate all reachable output graphs to a given input graphs, the program computation may not succeed. This is because GP(2)'s non-deterministic rule application operators are not guaranteed to find these reachable output graphs and programs may diverge~\cite{Plum09-TheGraph}.

\color{black}
\textit{GROOVE} is a tool that focuses on using verification and state-space exploration of GT systems~\cite{Groove03}. Given a GT specification, the \textit{GROOVE} simulator is capable of recursively generating (a possibly infinite) graph transition system (GTS). Rule applications can be controlled by specifying priorities. Moreover, \textit{GROOVE} provides a simple textual control language, e.g., with operators for conditionals, loops and random choice. The tool offers different state space exploration strategies, including full exploration (depth-first or breadth-first) and partial exploration (linear, random linear, and condition). Partial exploration strategoes are useful in cases where the state space is too large.  The tool has a mechanism to detect graph state collisions during exploration (up to graph isomorphism)~\cite{rensink2007isomorphism}. That mechanism uses a hashing function to compute graph certificates for predicting isomorphic graphs efficiently and with high accuracy. The implementation of the GTA in our tool (\textit{GrapeVine}) uses the same approach to detect graphs that are likely isomorphic.     

\color{black}

AGG is another GT tool with a formally defined execution semantics~\cite{Tae04-AGG,AGG2.0}. However, AGG avoids backtracking altogether and uses a random selection when making a non-deterministic choices during rule application and matching. \textit{FUJABA} has control structures based on a combination of UML activity diagrams and GT rules~\cite{FUJABA}. The semantics has not been defined formally and there is no backtracking. \textit{GrGen} provides a textual language for controlling GT rule applications but also without a formal semantics and backtracking~\cite{grgen}. 

GReAT is a GT tool with dedicated support for model transformations~\cite{Agrawal_2006}. It is notably different from other tools in that it support arbitrarily many input and output graphs. Its control structures use a combination of control-flow diagrams, structuring constructs and OCL-based model constraints.   

\section{Programmed Graph Transformation Systems with GTA}
\label{sec:gta}
This section is divided in two parts. We begin by defining our notion of a programmed graph transformation system (GTS) and then discuss the rational  behind the choices we made.

\subsection{Syntax and Semantics}

\begin{definition}[(programmed) Graph Transformation System (GTS)]
We define a (programmed) \emph{Graph Transformation System} (GTS) as a tuple $(R,C,P)$, where  $R$ is a set of \emph{rules}, $C$ is a set of \emph{constraints}, and  $P$ is a set of graph {\emph programs}.
\end{definition}

In this paper, we are mainly concerned with defining a formal basis for the programs $P$ of a GTS. We define the \textit{graph transformation control algebra} (GTA) for that purpose. GTA operators work on a common data type, a sequence of graph sets, referred to as a \textit{{\bf \em grap}h set {\bf \em e}numeration} (or ``\emph{grape}" for short) in the rest of this paper. We define this data type below. Let $\mathbb{G}$ denote the domain of graphs.

\begin{definition}[Graph set enumeration (grape)]
A \emph{graph set enumeration} (or {\em grape} for short) is a finite, non-empty sequence $\dddot{G}:\langle \bar{G_1}, \bar{G_2}, .., \bar{G_n} \rangle$ where each $\bar{G_i}$ is a finite set of  (possibly isomorphic) \color{black} graphs, i.e., $\bar{G_i} \subset \mathbb{G}$. Let  $\dddot{\mathbb{G}}$ denote the domain (data type) of \emph{grapes}. \color{black} Furthermore, we define the constant $\medstar$ to denote the \textit{grape} with with a single element, containing a single, empty graph, i.e., $\medstar = \langle {((\varnothing,\varnothing,\varnothing,\varnothing), \varnothing)} \rangle $.  \color{black}
\end{definition}

The GTA has ten operators representing functions on \textit{grapes}. We first introduce their syntax and then proceed to defining their semantics. 

\begin{definition}[Graph Program (syntax)]
Given a GTS $(R,C,P)$, $c \in C$ and $ r \in R$, each program $p \in P$ is a GTA expression, which is defined as one of the following:
 \begin{itemize}
    \item $\trianglepa(c)$ and $\trianglepacross(c)$ are GTA expressions, called \emph{constrain} and \emph{unconstrain}, respectively ;
    
    \item $\twoheadrightarrow(r)$ is a GTA expressions, called \emph{derive};
    \item $\circledcirc(n,\lesssim)$ with $n \in \mathbb{N}$ and a total order $\lesssim$ on graphs is a GTA expression, called \emph{select};
    \item $\vertdiv(e_1, e_2)$ and $\div(e_1, e_2)$ are GTA expressions, if $e_1$ and $e_2$ are GTA expressions; They are called \emph{sequence} and \emph{alternative}, respectively;
    \item $\circlearrowright(e)$ is a GTA expression called \emph{loop} if $e$ is a GTA expression; 
        \item $\looparrowright(c,e)$ is a GTA expression called \emph{search} if $e$ is a GTA expression;
    \item $\separated$ and  $\nequiv$ are GTA expressions called \color{black} \emph{cut} \color{black} and \emph{distinct}, respectively. 
\end{itemize}
\end{definition}

\begin{definition}[Graph Program (semantics)]
A graph program $p$ is interpreted as a function mapping \emph{grapes} to \emph{grapes}, i.e., $\llbracket p\rrbracket: \dddot{\mathbb{G}} \rightarrow \dddot{\mathbb{G}}$. The program semantics is based on the interpretation of the individual GTA operators as functions  $\dddot{\mathbb{G}} \rightarrow \dddot{\mathbb{G}}$. Most  of them are interpreted as total functions, with the exception of \emph{loop} and \emph{search} ($\circlearrowright, \looparrowright$) which are not guaranteed to terminate.

\begin{itemize}

\item {\bf Constrain and Unconstrain ($\trianglepa$ and $\trianglepacross$)}\\
The purpose of these functions is to declare and undeclare graph constraints, respectively. 
 $\trianglepa(c)$ declares constraint $c$ on all the graphs in the last element of a given \emph{grape}  that satisfy $c$. All other graphs are removed from that element. Formally:  $\llbracket \trianglepa(c)\rrbracket(\langle  .., \bar{G_n} \rangle)=\langle  .., \bar{G'_n} \rangle $ with $  \bar{G'_n}=  \{(G,K+\{c\}) \, | \, (G,K) \in  \bar{G_n} \land G \models c\}$
 
\vspace{.2cm}

 $\trianglepacross(c)$ removes constraint $c$ from  the graphs in the last element of a grape:  $ \llbracket \trianglepacross(c)\rrbracket(\langle  .., \bar{G_n} \rangle)=   \langle  .., \bar{G'_n} \rangle $  with $\bar{G'_n}= \{(G,K-\{c\}) \, | \, (G,K) \in  \bar{G_n} \}$

\vspace{.2cm}

Please note that whenever we use the notation of $\langle ..,  \rangle$ on two sides of a definition (like above), we require that ``$..$" stands for the \textit{same} sequence of elements. We use the notation   ``$..a$" and  ``$..b$" to refer to \textit{different} sequences of elements.

\item{\bf Derive ($\twoheadrightarrow$) }\\
Function \emph{derive} computes all direct linear derivation of each graph in the last element of a given \emph{grape} and extends the given input \emph{grape} with an element that contains \emph{all} resulting graphs, i.e.,  \newline
$\llbracket \twoheadrightarrow(r)\rrbracket(\langle .., \bar{G_n} \rangle) =\langle .., \bar{G_n}, \bar{G}_{n+1} \rangle $ where $\bar{G}_{n+1} = \{ G' | \exists G \in \bar{G_n} : G \overset{r}{\rightsquigarrow} G' \}$

\item{\bf Select ($\circledcirc$)} \\
Function \emph{Select} ($\circledcirc(k,\lesssim)$) reduces the last element of a given \emph{grape} to at most $k$ elements. The selection is determined by a total order on graphs $\lesssim$. Formally,     
 $\llbracket \circledcirc(k,\lesssim)\rrbracket(\langle .., \bar{G_n} \rangle)= \langle .., \bar{G'_n} \rangle$, with $ \bar{G'_n} \subseteq \bar{G_n} \land |\bar{G'_n}| \leq k \land \color{black} (|\bar{G}'_n| < k \Rightarrow \bar{G}'_n = \bar{G}_n) \land \color{black} \nexists G \in \bar{G_n}-\bar{G'_n}, G' \in \bar{G'_n}: G' \lesssim G$

\item{\bf Sequence ($\vertdiv$)}\\ \color{black}
 $\vertdiv(a,b)$ composes two GTA expressions sequentially by using functional composition, i.e., $\llbracket \vertdiv(a,b)\rrbracket=\llbracket b \rrbracket \circ \llbracket a \rrbracket. $ \color{black}

\item{\bf Alternative ($\div$)}\\
 $\div(a,b)$ composes two GTA expressions ($a$ and $b$) as alternatives by extending a given \emph{grape} with a new element that is the union of the last elements of the \emph{grapes} produced by interpreting the two expressions, i.e., $ \llbracket \div(a,b)\rrbracket(\dddot{G}:\langle ..x \rangle) =
 \langle ..x, \bar{O}_{1} \cup \bar{O}_{2}  \rangle$ with  $\llbracket a\rrbracket(\dddot{G})=\langle ..y, \bar{O_1} \rangle$ and $\llbracket b\rrbracket(\dddot{G})=\langle ..z, \bar{O_2} \rangle$.

\item{\bf Distinct $\nequiv$} \\
Graph exploration may produce identical graphs (up to isomorphism). 
\color{black}
The \emph{distinct} function  ensures that the last element of the \emph{grape} contains only unique graphs (up to isomorphism) when considering all graphs in the \emph{grape}. If the last element of the \emph{grape} contains several identical graphs, the smallest one is preserved (according to a given total order $\lesssim$). 
Formally, $\nequiv(\lesssim)$ is defined as  

  $\llbracket \nequiv(\lesssim) \rrbracket(\langle\bar{G}_1, \ldots, \bar{G}_{n-1} , \bar{G}_n\rangle) = (\langle\bar{G}_1, \ldots,  \bar{G}_{n-1} ,\bar{G}'_n\rangle)$
where
$\bar{G}'_n = \bar{G}_n \setminus \{D \mid (\exists i<n: \exists J \in \bar{G}_i: J \cong D) \vee (\exists D' \in \bar{G}_n: D' \lesssim D \wedge D' \cong D)\}$
\color{black}

  \color{black} Note that we disregard any constraints attached to constrained graphs when comparing them for isomorphism, i.e., $(G_1,K_1) \cong (G_2,K_2) \Leftrightarrow G_1 \cong G_2 $ for any two constrained graphs $(G_1,K_1)$ and $ (G_2,K_2)$.  \color{black}

 \item{\bf \color{black} Cut ($\separated$) \color{black} }\\
 \color{black} The \emph{cut} operator ``cuts off" all  but the last element of a given \textit{grape}.  \color{black} This is useful to restrict the \textit{distinct}'s operators ability to ``look back" in the derivation history to find identical graphs.  \emph{cut} is defined as $ \llbracket \separated \rrbracket (\langle .., \bar{G_n} \rangle) = \langle \bar{G_n} \rangle$.

\item{\bf Loop ($\circlearrowright$)}\\
$ \circlearrowright(e)$ is interpreted as a function that recursively interprets GTA expression $e$ on the most recently computed \emph{grape}  while the last element is not empty, i.e,

\begin{equation*}
  \llbracket \circlearrowright(e) \rrbracket (\dddot{G})=
    \begin{cases}
        \dddot{G}, & \text{if} \,  \llbracket e \rrbracket(\dddot{G})= \langle .., \varnothing \rangle \\
       \llbracket \circlearrowright(e) \rrbracket \circ \llbracket e \rrbracket(\dddot{G}) & \text{otherwise}
    \end{cases}
  \end{equation*}

\item{\bf Search ($\looparrowright$)}\\
$ \looparrowright(c,o)$ is interpreted as a (recursive) function that repeatedly interprets a GTA expression $o$ on the most recently computed \emph{grape}  while none of the graphs in the last element of the current \emph{grape} satisfy constraint $c$ and the last element is not empty, i.e,

\begin{equation*}
  \llbracket \looparrowright(c,o) \rrbracket (\dddot{G}:\langle .., \bar{G_n} \rangle)=
    \begin{cases}
        \dddot{G}, & \text{if} \, \bar{G_n} = \varnothing \vee \exists G \in  \bar{G_n} : G \models c \\
       \llbracket \looparrowright(c,o) \rrbracket  \circ \llbracket o \rrbracket (\ddot{G}) & \text{otherwise}
    \end{cases}
  \end{equation*}

\end{itemize}

\end{definition}

\subsection{Properties of the GTA and Rationale}

Habel and Plump showed that any graph programming language capable of (nondeterministic) rule application, sequential composition, and iteration is computationally complete ~\cite{HP01:ComputationalCompleteness}. They also showed that this set of operators is minimal. While Habel and Plump's rule application operator ($\Rightarrow_R$) combines rule selection (from a set of alternatives $R$) and rule matching/derivation, the GTA presented in this paper has separate operations for choosing alternatives  ($\div$) and matching/derivation ($\twoheadrightarrow$). However, it is easy to see that the behaviour of the $\Rightarrow_R$  operator can be achieved by combining the two GTA operators. The fact that these  GTA operators are deterministic means that they can produce all possible outputs of the non-deterministic operator. Since the GTA also has operators for sequential composition ($\vertdiv$) and iteration ($\circlearrowright$), the set of operators $\{\twoheadrightarrow, \div, \vertdiv, \circlearrowright \}$ is computationally complete (and minimal). 

To reduce the ``state-space explosion" problem of breadth-first exploration, the GTA has operators for reducing the search space. In particular, \color{black} the \textit{Distinct} operator ($\nequiv$) was introduced to detect and remove equivalent graph states (up to isomorphism). \color{black} This operator is the main reason to consider \textit{grapes} rather than simple graph sets as the common data type for computation.  The {\em Select} operator ($\circledcirc$) provides a way to limit the search to an upper number of ranked graphs to explore (based on a given total order). \color{black} The order relation used  by the \textit{Select} operator may be defined by the user according to the specifics of the graph program application. For example, it may be based on the notion of a graph-edit distance (GED) that measures the distance of the computed graphs from an objective target graph~\cite{10.5555/2736754.2736796}. While the general problem of computing GEDs is NP-hard, algorithms for efficient approximations exist.  \color{black}
For similar reasons (i.e., to restrict the search space), we included graph constraints in addition to GT rules. \textit{Constrain} ($\trianglepa$) and \textit{Unconstrain} ($\trianglepacross$) expressions provide programs with ways to (temporarily) restrict the exploration of possible computation.

\section{Implementing the GTA within \emph{GrapeVine}}
\label{sec-impl}

An implementation of GTA-based graph programs requires a functional GT tool in which graphs are immutable first-class objects rather than treating \textit{the} graph as ``singleton" global variable for stateful manipulation. We have recently proposed \textit{GrapeVine} as a tool to meet this criterion~\cite{Web22-ToolSupport}. We start this section with a short introduction to this tool, before discussing graph programs in \textit{GrapeVine}.

\subsection{\emph{GrapeVine}: A Short Overview}

\textit{GrapeVine} is implemented on top of the Neo4J graph database and derives much of its scalability from that architecture. The tool is a fundamentally new revision of the earlier tools \emph{Grape}~\cite{Web17-GrapeA} and \emph{Grape Press}~\cite{Web21-GrapePress}. Compared to these earlier tools, the main novelty in \textit{GrapeVine} is its functional computational model, where graphs are considered immutable objects and programs can be deterministic.

\textit{GrapeVine} uses a textual language for defining GT rules, constraints and programs. That language is provided as an internal DSL (domain-specific language) to the general-purpose programming language Clojure. \textit{GrapeVine} has been integrated with \textit{Gorilla}\footnote{http://gorilla-repl.org}. Gorilla is a browser-based  computational notebook for Clojure, similar to other computational notebook technologies, like the well-known Jupyter notebooks. Using this platform, \textit{GrapeVine} supports visualization of graphs, rules, and constraints (among other things)~\cite{Web21-GrapePress}. \textit{GrapeVine} can also be used independently of the computational notebook UI; \textit{GrapeVine} programs are regular Clojure programs, which run on the Java Virtual Machine (JVM) and can thus integrate with any other JVM language, such as Java, Kotlin, Scala, Groovy, etc.

\subsection{\emph{GrapeVine} Control Structures}
When designing the control structures for \textit{GrapeVine} we used the constructs provided by the functional host language (Clojure) as much as possible and avoided the creation of unnecessary ``syntactic sugar". For example, rule applications (derivations) can simply be called by the name of the rule, i.e., declaring a rule creates rule application functions with that name. Similarly, declaring a constraint automatically creates a pair of functions to check that constraint (positively and negatively).

Tab.~\ref{table:1} provides a correspondence between the  syntax used in {\em GrapeVine} graph programs and their interpretation in the form of GTA expressions.  As mentioned above, a rule application is simply a call to a function with the rule's name. Similarly, constraint checks are done by calling a constraint by name (with an added minus sign for negation). 
Constraints are added to graphs or removed from graphs using the \texttt{schema/schema-drop} functions. \textit{GrapeVine} uses the regular Clojure threading macro (\texttt{->}) for sequential composition and introduce a new macro (\text{||}) for composing alternatives.  In addition to a simple ``While possible" loop, \textit{GrapeVine} has two variations of an ``Until" loop. The behaviour of these variants is similar but the \texttt{->?+} operator (``Distinct Until") ensured that the computed graphs are unique. 

\begin{table}[h!]
\centering
\begin{tabular}{ |c|c|c| } 
 \hline
 Description &  Concrete Syntax (\textit{GrapeVine})& Interpretation (GTA) \\ 
 \hline
 \hline
 Application of rule \textit{r} & \texttt{r} & $\twoheadrightarrow(r)$ \\
\hline 
  Addition of  constraint $c$  & \texttt{schema c ..}  & $\trianglepa(c)$ \\ 
   (negated) & \texttt{schema c- ..}  & $\trianglepa(\neg c)$ \\ \hline
     Removal of  constraint $c$ & \texttt{schema-drop c ..} & $\trianglepacross(c)$  \\ 
     (negated) & \texttt{schema-drop c- ..} & $\trianglepacross(\neg c)$ \\ \hline
 Check of constraint $c$  & \texttt{c}  & $ \vertdiv(\trianglepa(c), \trianglepacross(c))$ \\
    (negated)  & \texttt{c-} & $\vertdiv(\trianglepa(\neg c), \trianglepacross(\neg c))$  \\ \hline

    Sequence  & \texttt{-> e\_1 e\_2 ..} & $\vertdiv(e_1,  \vertdiv(e_2, ..))$ \\ \hline
        Alternative & \texttt{|| e\_1 e\_2 ..} & $\div(e_1, \div(e_2, ..))$  \\ \hline
                 
   Loop (while possible)  & \texttt{->* e ..}  & $\circlearrowright(\vertdiv(e,..))$\\ \hline
   Until & \texttt{->?* c e ..} & $\looparrowright(c,\vertdiv(e,..))$  \\ \hline
      Distinct Until  & \texttt{->?+ c e ..}  & $\looparrowright(c,\vertdiv(\vertdiv(e,..),\nequiv)) \circ \separated$ \\ \hline
      Cut  & \texttt{cut}  & $\separated$\\ \hline
        A \textit{grape} with a single element       & \texttt{newgrape}  & $\medstar$\\ 
             containing an empty graph   & & \\ \hline
      Distinct   & \texttt{dist}  & $\nequiv$ \\ \hline
      Select & \texttt{select k v}   & $\circledcirc(k, v)$  \\ \hline

\end{tabular}
\caption{Concrete syntax of \textit{GrapeVine} control structures and their GTA interpretations}
\label{table:1}
\end{table}

Finally, \textit{GrapeVine} provides functions \texttt{cut}, \texttt{newgrape}, \texttt{dist}, and \texttt{select}, which directly correspond to the GTA operations $\separated$, $\medstar$, $\nequiv$, and $\circledcirc$.

Note that some \textit{GrapeVine} control structures are variadic, i.e., they allow an arbitrary number of parameters. This is indicated by  the ellipses in middle column of  Tab.~\ref{table:1}. \color{black} Variadic parameter lists are interpreted in the obvious way for adding/dropping multiple graph constraints and creating sequences/alternatives over multiple operations. In the case of loops, longer parameter lists are interpreted as sequential composition blocks of operations that are performed at each iteration. \color{black}

\subsection{\emph{GrapeVine} Control Structures: A Simple Example}
Let us look at a simple example of a graph program in \textit{GrapeVine} to get an impression. We use the example of the well-known ferryman problem~\cite{ZS91-NondeterministicControl}. In that problem, a ferryman needs to (safely) transport three things (a goat, a wolf, and a cabbage) from one side of a river to the other side. He can only transport one thing at a time.  The wolf will eat the goat and the goat will eat the cabbage, respectively, if left unattended with their food. 

\begin{figure}
    \centering
    \includegraphics[width=16cm]{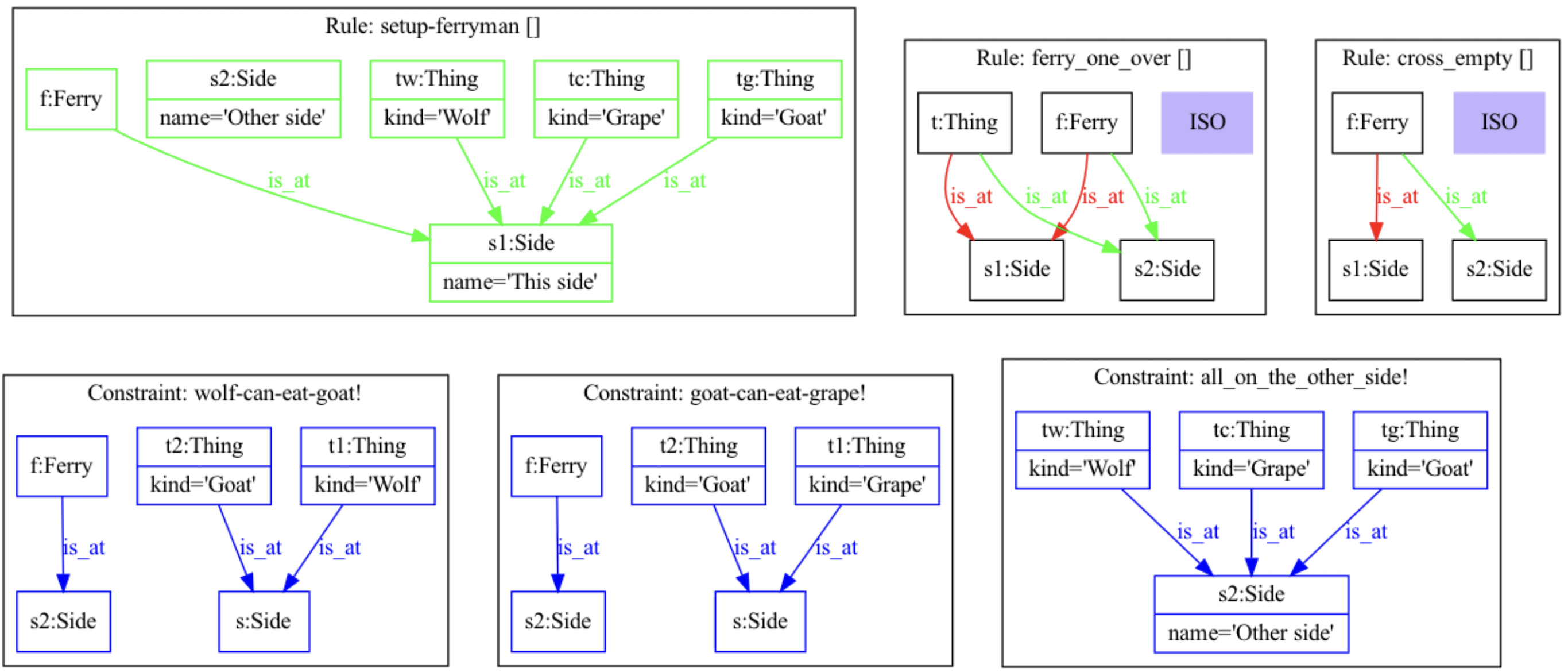}
    \caption{Ferryman problem (rules and constraints)}
    \label{fig:fman}
\end{figure}

Fig.\ref{fig:fman} shows three rules and three constraints modelled within \textit{GrapeVine} in support of solving the ferryman problem. (For simplicity, we omit the textual definition of these rules and only show their generated graphical view. Also, please note that  for artistic freedom, the cabbage was exchanged by a grape.) Rule \textit{setup-ferryman} creates the initial graph with the two river sides and all items on one side. Created graph elements are shown in green. The other two rules \textit{ferry\_one\_over} and \textit{cross\_empty} specify the two possible actions, namely to cross with or without an item. Deleted graph elements are red. 

The shaded isomorphism (``\texttt{ISO}") box in the two rules specify that these rules must use \textit{isomorphic} subgraph matching. (By default, \textit{GrapeVine} does not enforce isomorphic matches, i.e.,  separate elements on a rule's left-hand side do not need to match to separate elements in the host graph.)

The bottom half of Fig.\ref{fig:fman} shows three (atomic) graph constraints. In \textit{GrapeVine}, basic graph constraints $I \xrightarrow{c} T$ are visualized by using black colour for elements in $I$ (``if" pattern) and blue colour for elements in $T$ (``then" pattern). In the case of this example, the three graph constraints have the form  $\varnothing \xrightarrow{c} T$, i.e., they represent simple existence constraints (of form $\exists T$). Such constraints are also called \textit{basic constraints}~\cite{OEP08-ALogic}. Note that pattern matching for constraints always uses isomorphic matching.

To better distinguish between the names of rules and constraints in graph programs, it is a convention in \textit{GrapeVine} to append an exclamation mark to the names of constraints. 

Fig.\ref{fig-gp1} shows a \textit{GrapeVine} program for solving the ferryman problem. The program should be understandable based on the above definition of control structures and the definition of the rules and constraints. The call to \texttt{newgrape} creates a new \textit{grape} with an empty graph. After setting up the initial state (applying rule \texttt{setup-ferryman}), we enter a loop that recurs until one of the graphs in the last element of the current \textit{grape} satisfies the constraint \texttt{all\_on\_the\_other\_side!}. 

Note that the program in Fig.~\ref{fig-gp1} is guaranteed to find a solution if there exists one. If we had used non-deterministic rule application and a depth-first exploration strategy, the execution mechanism may or may not find a solution, depending on how it was implemented. A mechanism that resolves non-deterministic choices by a fair process (e.g., by using true randomness) will eventually find a solution, but without bounded time. 

\begin{figure}[h]
\begin{verbatim}
(-> (newgrape) setup-ferryman
    (->?+ all_on_the_other_side!
          (|| ferry_one_over cross_empty)
          wolf-can-eat-goat!-
          goat-can-eat-grape!-)))
\end{verbatim}
\caption{A program to solve the ferryman problem}
\label{fig-gp1}
\end{figure}

Fig.~\ref{fig-gp2} shows an alternative, yet equivalent program for solving the ferryman problem. This program declares the two constraints we checked manually in Fig.~\ref{fig-gp1} as invariant constraints on the graphs in the \textit{grape}. Once declared, \textit{GrapeVine} will ensure that only valid derivations are possible.

\begin{figure}[h]
\begin{verbatim}
(-> (newgrape) 
    (schema wolf-can-eat-goat!- goat-can-eat-grape!-) 
    setup-ferryman
    (->?+ all_on_the_other_side 
          (|| ferry_one_over cross_empty)))
\end{verbatim}
\caption{An alternative program, using schema constraints}
\label{fig-gp2}
\end{figure}

\color{black}
\subsection{Example Execution}

\begin{figure}
    \centering
    \includegraphics[width=16cm]{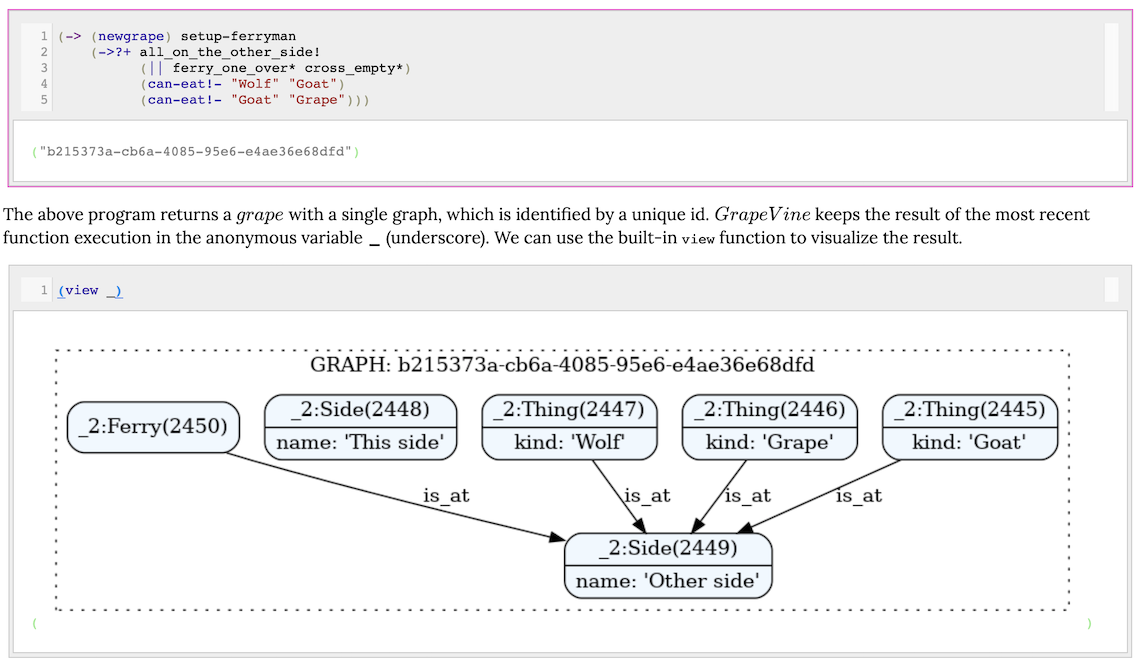}
    \caption{Sample execution of Ferryman program}
    \label{fig:exec1}
\end{figure}

Fig.~\ref{fig:exec1} shows a screen capture of an example execution of the Ferryman program, using \textit{GrapeVine's} computational notebook UI. The figure shows two executable segments (boxes) where the shaded part contains executable code and the white part shows the output of the execution. The upper code segment shows that executing our Ferryman program returns a collection with a single element.  (Like all Lisp-like languages, Clojure uses parentheses to indicate lists.) The returned list represents the last element of the \textit{grape} that is returned by calling the functional graph program. Each graph is referred to by a unique ID. In this case, there is only one single graph.

Graphs can be visualized in the computational notebook using the notebook's \textit{view} function. This is shown in the second executable segment in Fig.~\ref{fig:exec1}. Note that for convenience, \textit{GrapeVine} provides a special variable ``\texttt{\_}" (underscore) to refer to the most recently computed \textit{grape}. 

\begin{figure}
    \centering
    \includegraphics[width=16cm]{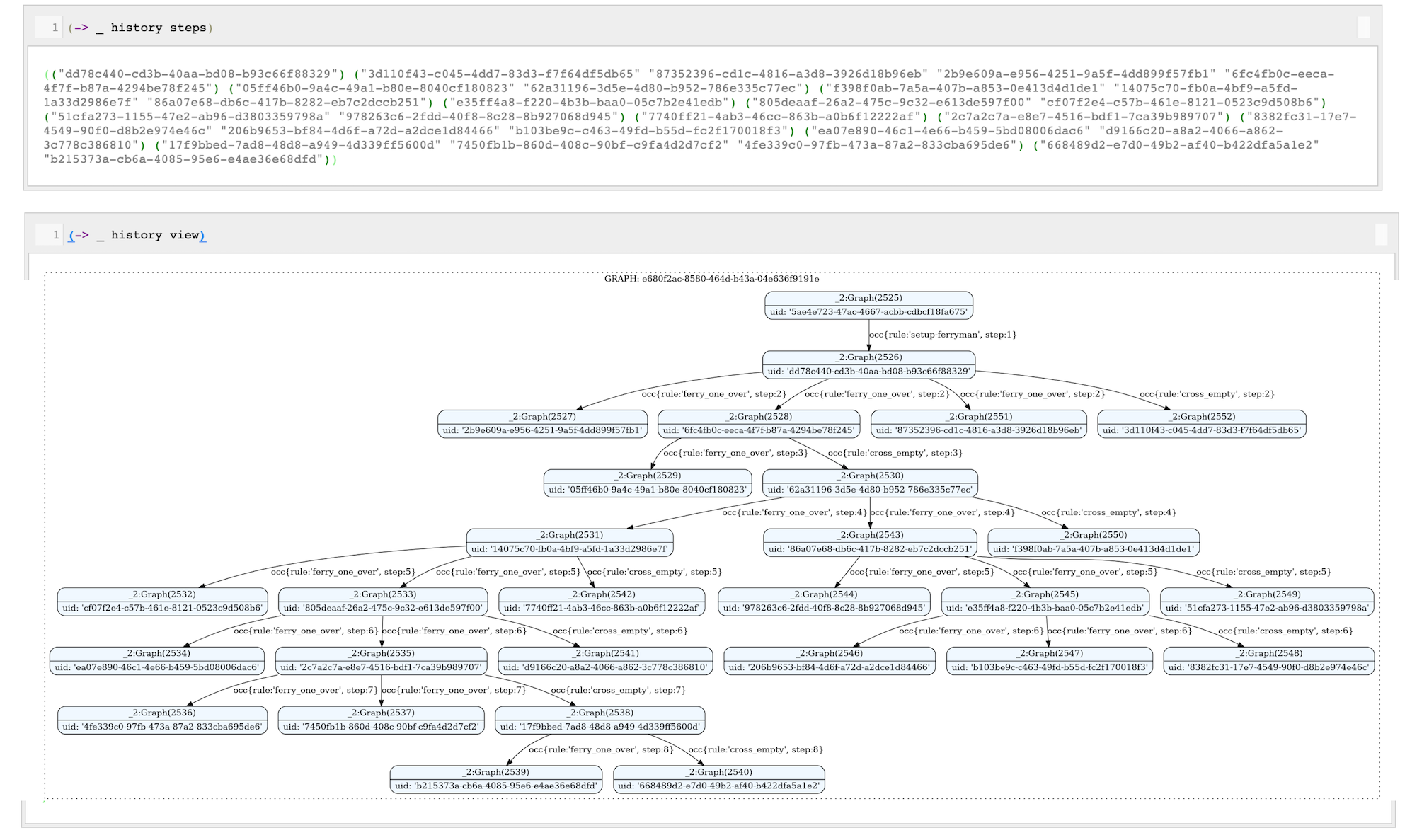}
    \caption{Visualizing the entire \textit{grape} (history) of the example execution}
    \label{fig:exec2}
\end{figure}

The reason for returning merely the last element of a \textit{grape} rather than the entire (history) sequence is for performance reasons and to avoid representational clutter in the computational notebook. Most of the times, we are only interested in the last element of a grape when computing with functional graph transformations. However, the full grape (history) is kept internally and can be returned when needed. This is demonstrated in Fig.~\ref{fig:exec2}, where the first code block shows the use of the built-in functions \texttt{history} and \texttt{steps} to output the entire \textit{grape} for our example computation (which is represented as a list of lists of graph IDs in Clojure). 

The textual output of a full \textit{grape} history is not easy to read. \textit{GrapeVine's} computational notebook UI has functions to visualize the history of a \textit{grape} by means of a graph of graphs, where each edge marks an occurrence of a rule application. This is demonstrated in the second executable segment in Fig.~\ref{fig:exec2}, where each ``row" of the shown ``history graph" represents a new element in the \textit{grape} sequence.

\begin{figure}
    \centering
    \includegraphics[width=16cm]{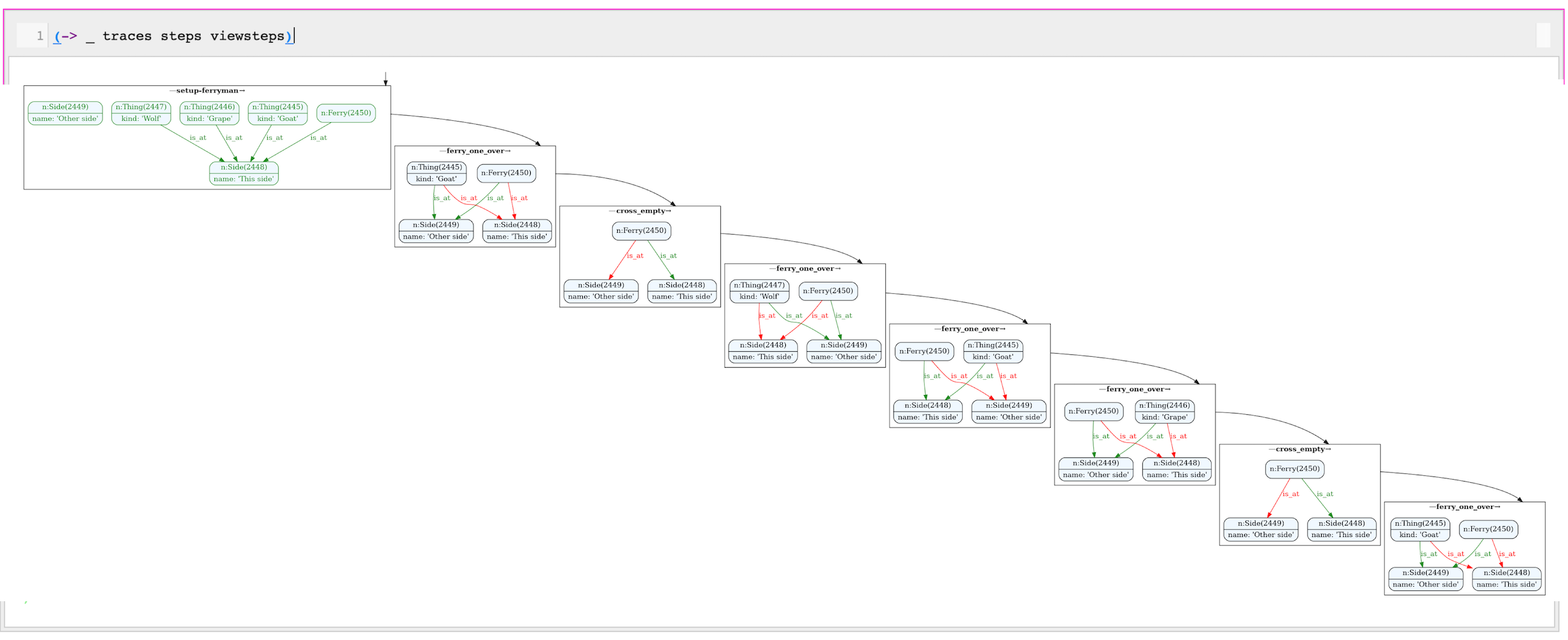}
    \caption{Visualizing the detailed steps for each rule occurrence in an execution trace}
    \label{fig:exec3}
\end{figure}

\textit{GrapeVine} also provides ways to visualize the details of each rule occurrence. Fig.~\ref{fig:exec3} demonstrates the use of the built-in \texttt{traces} function to visualize the details of rule occurrences in the history of computing the result of our Ferryman solution. 

\color{black}

\subsection{Efficiency Considerations}

An important factor for making the above described approach to GT programming feasible is the ability to avoid exploring graph states that have been explored before (by means of the $\nequiv$ operator). This operator requires that previous versions of graphs are efficiently accessible (and comparable). Moreover, parallel exploration of possible choices generate non-linear version histories of graphs that must be maintained efficiently. \textit{GrapeVine} uses a fully-persistent data structure for maintaining graphs in a graph database (Neo4J). Details on that persistent data structure have been reported in \cite{Web22-ToolSupport} and interested readers are referred to that publication. However, one new aspect that we want to comment on is the implementation of the $\nequiv$ operator. While graph isomorphism checks are expensive, Rensink has demonstrated that hashed graph ``certificates" can be used to speed up similarity checking in practice  based on the idea of bisimulation~\cite{rensink2007isomorphism}.   \textit{GrapeVine} uses this approach for comparing graphs. Moreover, since graphs are immutable, their hash certificates are computed only once, at creation time and then stored in the database using an efficient indexed search structure. \color{black} This means that the run-time for similarity checks only grows logarithmically with the number of graphs in the \textit{grape}, given the search and insertion complexity of b+ trees. \color{black}

When compared to some other GT tools, specifically those that perform all computation in main memory, \color{black}  \textit{GrapeVine} has a much higher latency when performing graph transformations. This is due to its client-server architecture between the Clojure (JVM) client and the database server. These two components communicate using a Web-service API and each rule application requires multiple interactions (e.g., for rule matching, checking of application conditions, rule application, checking of graph constraints). However, \textit{GrapeVine's} performance is scalable with respect to the number and size of graphs.  \color{black}
While a comprehensive performance analysis is out of the scope of this paper, we have conducted a preliminary experiment with the program in Fig.~\ref{fig:fman}. The platform was a 2017 iMac Pro with a 3Ghz Intel Xeon processor running both the database server (Neo4J) and \textit{GrapeVine} in separate Docker containers connected by a bridged network. 

Running the Ferryman program  in Fig.~\ref{fig:fman} on an empty graph database takes approximately 7 seconds and creates 27 graphs in the process. The execution time of the program remains the same (approximately 7 seconds) even after running the program 1000 times and creating 27,000 graphs in the database. (Note: \textit{GrapeVine} does not automatically ``garbage collect" graphs, but the user can trigger a garbage collector manually~\cite{Web22-ToolSupport}.)

If we modify the program in Fig.~\ref{fig:fman} to use the ``Until" operator without the distinction check (\texttt{->?*} operator instead of \texttt{->?+}), the program creates 216 graphs and takes 52 seconds. 

We can compare these measurements with \textit{GrapeVine's} predecessor (\textit{Grape}), which implements a backtracking-based depth-first exploration using non-deterministic GT operators. This comparison makes sense, since it has the same Web-service based architecture (JVM client and Neo4J database server.) Client-server interactions therefore have similar latency. \textit{Grape} is actually not able to compute a solution for the ferryman problem as stated in Fig.~\ref{fig:fman}. This is because its execution engine is deterministic and would  always explore non-deterministic alternatives between applicable rules in the order they are stated in the program. In the case of our program in Fig.~\ref{fig:fman}, this would result in a diverging program run that would take the goat back and forth forever. Note that this problem would not go away if the order of the ``alternative" statement was swapped. In that case, the goat would be moved and the ferryman would travel empty forever.  \cite{Web17-GrapeA} provides a modified program that uses bounded depth-first search (by giving the ferryman a maximum budget of seven moves). That program takes approximately 50 seconds to find a solution on the same machine as above. Of course, we note that \textit{Grape} does not check for collisions when exploring solutions depth-first. Doing so would certainly increase efficiency. However, the comparison shows that the deterministic (breadth-first) operators in \textit{GrapeVine} have a performance that is competitive with  traditional (depth-first, backtracking-based) solutions to resolve non-determinism.

\section{Conclusions and Current Work}
\label{sec-conc}
While graph transformations have a well-defined semantics, the control structures used for programming with GTs are often either of limited expressiveness (but formalized) or they are expressive but lack a formal semantics. Implementing non-deterministic control structures in GT tools that use stateful computation requires complex and often inefficient processing such as backtracking on graph exploration. Unbounded depth-first search strategies may not find solutions that are theoretically computable. For example, a solution for the above-mentioned ferryman problem may not be found in GT tools that implement non-deterministic rule application based on depth-first search and backtracking unless repeated exploration of the same graph states are avoided or randomized exploration strategies are used. 

Functional GT tools avoid many of the complexities of stateful GT tools, since graphs are immutable (i.e., derivations do not destroy/modify  the input graph) and rule application can be deterministic. In this paper, we have defined a formal foundation for control structures used for functional GT, called the {\em Graph Transformation control Algebra} (GTA). GTA-based programs can be written as deterministic functions, i.e., they guarantee that all graphs that can be derived via a particular program can be returned. To make this feasible, GTA operators provide means to limit the exploration of the search space, for example by detecting previously explored graphs, checking graph constraints and selecting graphs based on a total ordering heuristic. 

We have shown the implementation of GTA-based control structures within the functional GT tool \textit{GrapeVine}. Our current work is on performing a more thorough performance evaluation with larger application problems and community benchmarks to provide further evidence for the feasibility of the proposed approach. \color{black} For an experience report of using \textit{GrapeVine}  for one such larger problem, we refer the readers to a thesis by Machowczyk, who applied the approach to an application of graph rewriting to graph neural networks at the University of Leicester~\cite{Ma22-GraphRewriting}. \color{black}

\nocite{*}
\bibliographystyle{eptcs}
\bibliography{generic}
\end{document}